\documentclass[letterpaper,twocolumn,10pt]{article}
\usepackage{usenix}

\usepackage{color}
\definecolor{green}{rgb}{0.0, 0.0, 0.00} 
\definecolor{orange}{rgb}{0.8, 0.6, 0.2}
\definecolor{red}{rgb}{0.88, 0.33, 0.33}
\definecolor{teal}{rgb}{0.2, 0.54, 0.65}
\definecolor{blue}{rgb}{0.3, 0.52, 0.88} 
\definecolor{purple}{rgb}{0.47,0.41, 0.68}
\definecolor{saffron}{rgb}{0.88, 0.74, 0.16}
\definecolor{turquoise}{rgb}{0.0, 0.5, 0.5}
\definecolor{black}{rgb}{0.0, 0.0, 0.0}

\newcommand{\todo}[1]{\marginpar{\Large{\color{red} $\spadesuit$}} {\bf \color{red} [#1]}}

\usepackage{amsmath,amsfonts,bm}

\def\figref#1{Fig.~\ref{fig:#1}}

\def\eqref#1{Eq.~(\ref{eq:#1})}

\def\secref#1{Sec.~\ref{sec:#1}}

\newcommand{\appref}[1]{Appx.~\ref{app:#1}}
\newcommand{\tabref}[1]{Table~\ref{tab:#1}}

\def\1{\bm{1}}

\DeclareMathAlphabet{\mathsfit}{\encodingdefault}{\sfdefault}{m}{sl}
\SetMathAlphabet{\mathsfit}{bold}{\encodingdefault}{\sfdefault}{bx}{n}

\newcommand{\secprespace}{\vspace{-0.9mm}} 
\newcommand{\secpostspace}{\vspace{-0.9mm}} 
\newcommand{\paraspace}{\vspace{-1.6mm}} 

\usepackage{hyperref}
\usepackage{url}

\usepackage{comment}
\usepackage{verbatim}
\usepackage{graphicx}
\usepackage{booktabs}
\usepackage{enumitem}
\usepackage{mathtools}
\usepackage[normalem]{ulem}
\usepackage{array,multirow}
\usepackage{float}
\usepackage{wrapfig}
\usepackage{cuted}

\microtypecontext{spacing=nonfrench}

\title{\Large \bf Can one hear the shape of a neural network?: 
\\ Snooping the GPU via Magnetic Side Channel}

\newcommand{\Li}{\textbf{(i)}\ }
\newcommand{\Lii}{\textbf{(ii)}\ }
\newcommand{\Liii}{\textbf{(iii)}\ }
\newcommand{\Liv}{\textbf{(iv)}\ }
\newcommand{\Lv}{\textbf{(v)}\ }

\date{}

\begin{document}

\author{
{\rm Henrique Teles Maia}\\
Columbia University
\and
{\rm Chang Xiao}\\
Columbia University
\and
{\rm Dingzeyu Li}\\
Adobe Research
\and
{\rm Eitan Grinspun}\\
Columbia University \& \\
University of Toronto
\and
{\rm Changxi Zheng}\\
Columbia University
} 

\maketitle

\begin{abstract}
Neural network applications have become popular
in both enterprise and personal settings.
Network solutions are tuned meticulously for each task, and designs
that can robustly resolve queries end up in high demand.
As the commercial value of accurate and performant machine learning models increases,
so too does the demand to protect neural architectures as confidential investments.
We explore the vulnerability of neural networks deployed as
black boxes across accelerated hardware through electromagnetic side channels.

We examine the magnetic flux emanating from
a graphics processing unit's 
power cable, as acquired by a cheap \$3 induction sensor, 
and find that this signal betrays the detailed topology and
hyperparameters of a black-box neural network model. 
The attack acquires the magnetic signal for one query 
with unknown input values, but known input dimensions.
The network reconstruction is possible due to the
modular layer sequence in which deep neural networks are evaluated.
We find that each layer component's evaluation
produces an identifiable magnetic signal signature, 
from which layer topology, width, function type, 
and sequence order can be inferred using a suitably trained 
classifier and a joint consistency optimization based on integer programming. 

We study the extent to which network specifications
can be recovered, and consider metrics for comparing network similarity.
We demonstrate the potential accuracy of this side channel attack
in recovering the details for a broad range of 
network architectures, including random designs.
We consider applications that may exploit this novel side channel exposure,
such as adversarial transfer attacks.
In response, we discuss countermeasures to 
protect against our method and other similar snooping techniques.


\end{abstract}

\secprespace
\section{Introduction}
\secpostspace

The graphics processing unit (GPU) is a favored vehicle for executing a neural network. 
GPUs allow difficult and sizable jobs to be treated faster, 
and have been used extensively in state of the art machine learning
pipelines across both academic and commercial settings.
In turn, the widespread success of neural network models when applied
to real-world challenges in vision~\cite{he2016vision},
security~\cite{xu2017security},
natural language processing~\cite{teufl2010nlp}, 
and robotics~\cite{kober2013robotics},
has solidified the demand for GPUs to support machine learning technologies.
Hardware and software companies alike have streamlined management of the 
GPU into their products as they strive to
support ever larger and more complex neural networks~\cite{nvidiaTriton,abadi2016tensorflow}.

These recent developments raise security concerns surrounding an adversary 
who wishes to uncover the underlying network design from an application. 
Model extraction attacks, aimed at reverse engineering a network structure,
have attracted a growing research
effort~\cite{joon2018towards,tramer2016stealing,wang2018stealing},
and are motivated by several incentives.
First, it is well
known that the performance of a network model hinges on its judiciously
designed structure---but finding an effective design is no easy task. 
Significant time and energy is expended in searching and fine-tuning network
structures~\cite{zoph2018learning}. 
Moreover, in industry, optimized network
structures are often considered confidential intellectual property. 
In some cases businesses even charge per 
inference for queries submitted by a client to networks
they host or provide as a 
service~\cite{amazonECSprice,googleaiprice}.

Furthermore, a reverse engineered ``surrogate'' model also makes
the black-box ``victim'' model more 
susceptible to adversarial \emph{transfer attacks}~\cite{papernot2017practical,liu2016delving}, in
which a vulnerability identified in the surrogate is exploited
on the victim. Success in the exploit is contingent
on the ability of the surrogate to successfully model the vulnerabilities of the victim.
Recovering accurate, detailed network topology and hyperparameters informs 
the modeling of a good surrogate.
It is therefore important to understand how this valuable, privileged
information can be compromised. 

These risks are assessed in the study of physical side-channel attacks targeting neural networks.
Granted local access to processors, several works have explored how 
electromagnetic (EM) radiation is a powerful
side channel through which one can infer network details
\cite{yu2020deepem,batina2019csi}.
However, these works make use of methods specified to a limited set of neural architectures
that are tailored for microprocessors and edge devices.
We apply their same threat model, but instead generalize our approach to handle
a wider array of deep neural networks running on the GPU.

Most similar to our goal is a recent effort that likewise targets models on a GPU
but by observing read-write volumes and memory traces via 
both EM signals and bus snooping~\cite{hu2020deepsniffer}.
However, rather than aiming to predict precise parameter dimensions 
for their extracted layers, their approach
randomly selects parameter values from predetermined sets.
We instead examine an alternative source of EM leakage, the power
supplied to the GPU, and develop a novel way to 
assign parameters for arbitrary layers that result in valid architectures.
Our work 
extends the capability
of previous EM side channel attacks
in order to generalize extraction for complex models running on advanced hardware.

\paraspace
\vspace{-1mm}
\paragraph{Approach.}
The GPU consumes energy at a variable rate that depends on operations performed. 
Every microprocessor instruction is driven by transistor electron flows, 
and different instructions require different power levels~\cite{grochowski2006energy}.  
The many compute cores of a GPU amplify the fluctuation in energy consumption, 
and so too the current drawn from the power cable. 
Current induces magnetic flux governed by the Biot-Savart law~\cite{griffiths2005introduction}, 
and current fluctuations induce EM ripples whose
propagation through the environment is governed by the Amp\`{e}re-Maxwell law.
Even a cheap, \$3 magnetic induction sensor (see \figref{physicalSensor}) 
placed within a few millimeters of the power cable suffices to record these EM ripples.

We examine the fluctuation of magnetic flux from the GPU's power cable,
and ask whether a {non-intrusive} observer can glean the information needed to 
reconstruct neural network structures.
Our findings span across multiple GPU models and 
demonstrate transfer attacks on state of the art networks.
Remarkably, we show that, through magnetic induction sensing,
a {passive} observer can reconstruct the \emph{complete} network structure even for 
\emph{large} and \emph{deep} networks.

To reconstruct the black-box network's structure, we propose a two-step approach.
First, we estimate the network topology, such as the number and types of layers, 
and types of activation functions,
using a suitably trained neural network classifier. 
Then, for each layer, we estimate its hyperparameters 
using another set of deep neural network (DNN) models.
The individually estimated hyperparameters are then jointly optimized by solving an 
integer programming problem to enforce consistency between the layers.
We demonstrate the potential accuracy of this side-channel attack in recovering 
the details for a wide range of networks, including large, 
deep networks such as ResNet101~\cite{he2016vision}.
We further apply this recovery approach to demonstrate black-box adversarial transfer attacks.

To summarize, our main contributions are as follows:
\begin{itemize}
    \item We study how the large-amplitude EM radiation associated to neural network GPU implementations
    allows recovery of significantly richer network details.
    \item We explore a simple--yet--effective recurrent classification model that translates measurements made by a cheap and elementary sensor.
    \item We present a robust algorithm based on limited a priori knowledge of parameters and heuristics, complete with an integer programming optimization formulation, which improves on previous attempts to specify network models from side-channel information.
    \item We demonstrate vulnerabilities through model extraction and transfer attack results that leverage the proposed in-depth recovery of a black box network model.
\end{itemize}

\secprespace
\section{Related Work}\label{sec:related}
\secpostspace
Our work falls under the umbrella of black-box model extraction.
Absent access to the model's internals, one might infer  
structure from observed input-output pairs. For instance, 
\cite{tramer2016stealing} demonstrated that,
for simple models such as decision trees and support vector machines hosted on a cloud,
certain internal information can be extracted via a large amount of queries.
This approach, which was extended to infer details of deep neural
networks~\cite{joon2018towards,liu2016delving,duddu2019quantifying,wang2018stealing}, 
is typically able to recover some crucial information, such as the optimization
learning rate and narrowing in on the network structure family, but
has not demonstrated recovery of full structural details.

\begin{figure*}[ht]
    \centering
    \vspace{-3mm}
    \includegraphics[width=0.95\textwidth]{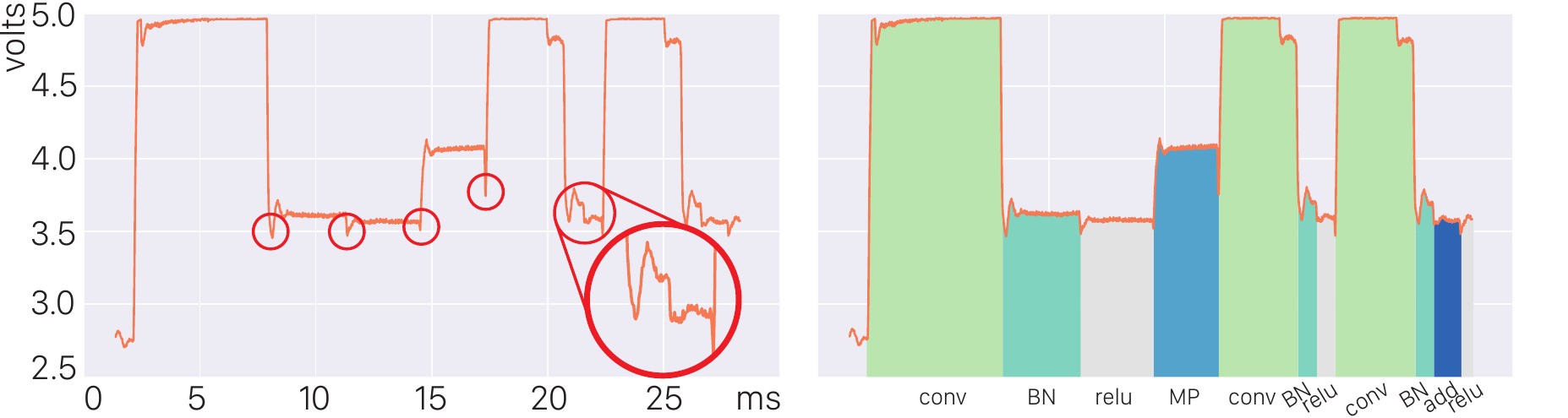}
    \vspace{-1mm}
    \caption{\textbf{Leaked magnetic signal.} 
    (left) Our induction sensor captures a magnetic signal when 
    a CNN is running on the GPU.
    We observe strong correlation between the signal and the network steps.
    Across two steps, the GPU has to synchronize, resulting in 
    a sharp drop of the signal level (highlighted by selected red circles).
    (right) We can accurately classify the network steps and reconstruct the topology, as indicated by the labels under the $x$-axis. Here we highlight the signal regions associated with convolutions (conv), batch-norm (BN), Relu non-linear activations (relu), max-pooling (MP), and adding steps together (add).
    \label{fig:signalClass}}
    \vspace{-2mm}
\end{figure*}

An orthogonal approach, side-channel analysis (SCA),
extracts information gained from the \emph{physical} implementation of a model, 
rather than in the mathematical model itself. 
Analysis of timing~\cite{kocher1996timing}, 
power~\cite{kocher1999differential,luo2015side}, 
cache flushes~\cite{yarom2014flush+}, 
and audio~\cite{genkin2014rsa} 
have been prominently demonstrated 
to extract secret keys
from cryptographic procedures such as the Digital Signature and Advanced Encryption
Standards.

When concerned with machine learning, different network models 
exert different computational burdens on hardware~\cite{xiang2020open}. 
The variance across operations and layers result in different physical
patterns of consumption, regardless of implementation or hardware, 
leaving neural architectures in general susceptible to side channel attacks.
SCA was recently used to infer machine learning models
by observing power consumption 
profiles~\cite{xiang2020open, wei2018know, dubeymaskednet}, 
timing information~\cite{duddu2018stealing} and memory/cache
access~\cite{hu2020deepsniffer,hong20190wn,hua2018reverse,yan2020cache}. 
These methods placed a malware process 
on the machine hosting the black-box model. 
Our threat model does not involve introducing processes on the host.

In cases where access to memory is secured by the target,
one can still leverage physical access to collect patterns 
from the processor itself or its power consumption.
Recently, side channel analysis of EM radiation
has been applied to exploit network model extraction \cite{batina2019csi, yu2020deepem}.
These attempts concentrate on EM radiation from embedded processors.
Edge devices are often constrained to \emph{lite} (abbreviated)
machine learning frameworks,
limiting the size and complexity of models supported by the hardware.
In contrast to GPUs, embedded processors emit a relatively weak EM signal,
necessitating delicate 
and costly 
measurement devices and even mechanical 
opening of the chip package.

\paragraph{Our advance.} 
Previous works are demonstrated on shallow networks
(e.g., fewer than 20 layers) and on more limited edge hardware. 
It remains unclear whether
these methods can also extract deep network models, 
ones that are structurally more complex 
and more prevalent in practice.
We demonstrate successful recovery of the full structure of deep networks,
such as AlexNet~\cite{krizhevsky2012imagenet}, 
VGGNet~\cite{Simonyan2014vgg}, 
and ResNet101~\cite{he2016vision}. 
With that, we hope to raise awareness of the GPU's EM radiation
as an information-rich, easily and non-intrusively probed side channel.

\secprespace
\section{Background}
\secpostspace

\subsection{Neural Networks}\label{sec:backNN}

Neural Networks constitute a field of machine learning algorithms suited
for general purpose feature extraction, regression, 
and classification objectives~\cite{lecun2015nature}.
A network is an assembly layers, each containing either a linear operation, 
an aggregate operator, or a non-linear activation function that works to transform an input.
Each layer consists of a differentiable function in order to permit optimization
of an objective through back-propagation and stochastic gradient descent.

The design of a neural network, often referred to as its architecture, model, or topology,
gives the network its shape.
Models are commonly defined by their shape characteristics, namely: 
\Li the number of layers used, also known as the depth,
\Lii the sequence in which layers appear, and
\Liii each layer's individual type 
(e.g. fully connected, convolutional, recurrent, pooling, activation, normalization, etc.).
Networks may then compose these elements in unique ways.
For example, VGGNet and ResNet are both composed of convolutional, 
pooling, activation, and fully connected layers, 
but structure them with distinct numbers of layers set up in differing orders.

Beyond the high-level shape of a network, 
a model must also specify the parameters for its layers.
This includes detailing the size and padding of convolution kernels,
whether to use average or max pooling,
or choice in activation function from a growing list of candidates (e.g., ReLU, Tanh, Sigmoid, etc.).
It also predominantly involves delineating the dimension for each layer.
In practice, a layer's size dimension 
determines the number of operations (such as multiplication or addition)
it imposes, and is closely related to the computational overhead of the layer.
Asides from transforming the values passing through the network,
size parameters also influence the dimensions of neighboring layers.
Layer definitions cannot be specified arbitrarily 
and must be consistent across layers in the network 
in order to ensure valid output-input dimension agreement.


After a neural architecture is chosen, networks undergo stages of training and inference.
Training leverages a labeled dataset of inputs and desired outputs to update values stored on
layers in order to 
minimize a target loss function.
Inferences make use of the trained network to resolve queries, leveraging patterns 
'learned' throughout training~\cite{lecun2015nature}.
A networks shape, along with the variance of the training set, dictates the
capability of a network to accurately and efficiently process new queries.

\subsection{GPUs for Deep Neural Networks}\label{sec:backGPU}



GPU hardware is used pervasively throughout the machine learning community,
and some GPUs even feature dedicated tensor-based cores 
optimized for computing network \emph{steps}~\cite{nvidiatensorcores}.
We use \emph{step} to refer to
performing a specific kind of network operation, 
such as a linear operation, batch normalization, pooling, activation function, etc. 
A \emph{layer} is a sequences of steps, 
e.g., a \Li linear operation, then \Lii pooling, then \Liii activation. 
While there may be data dependencies between steps,
there are no such dependencies within a step.

The parallel nature of GPU computation lends itself to a natural implementation
of networks, wherein each \emph{step} defines a compute kernel that is executed in parallel, 
i.e., single instruction multiple data (SIMD) parallelism. 
Transitions between steps, however, are synchronized~\cite{buck2007gpu}: in
our example above, activation begins only after pooling completes.
This cross-step synchronization allows for implementations structured into
modules, or GPU \emph{kernels}. This modular approach is
employed in widely-used deep learning frameworks 
such as {PyTorch} and {TensorFlow}~\cite{paszke2019pytorch,abadi2016tensorflow}.

\subsection{Magnetic Signals from GPUs}\label{sec:signal}

%




Kernel execution demands transistor flips, which
place electric load on the GPU processor,
in turn emitting magnetic flux from its power cable.
An induction sensor measures this flux
and produces proportional voltage. 
The time-varying voltage is our acquired \emph{signal}
(see \figref{signalClass}).

Different steps correspond to different GPU kernels, transistor flips,
electric loads, and signal characteristics, which are distinguished even
by the naked eye (see \figref{signalClass}). 
Cross-step synchronization involves idling, dramatically reducing electric
load and signal level (see \figref{signalClass}).
These rapid sharp drops demarcate steps.


We observe that the acquired signal strongly correlates to the \emph{kind} of GPU operations, 
rather than the specific \emph{values} of computed floating point numbers.
We verify this by examining signals using both PyTorch and TensorFlow
and on multiple GPU models (see \secref{ret}). 
Furthermore, we discuss how the signal is processed in~\secref{topo},
and later address challenges to our side-channel signal in~\secref{defense}.


The signal is also affected by the input to the network.
Although the specific input data values do not influence the signal,
the input data size does.
When the GPU launches a network, the size of its single input (e.g., image resolution) is fixed. 
But the network may be provided with a batch of input data (e.g., multiple images).
As the batch size increases, more GPU cores will be utilized in each step. The
GPU consequently draws more power, which in turn strengthens the
signal. 
Once all GPU cores are involved, further increase of input batch size  will not increase the 
signal strength, but elongate the execution time until the GPU runs out of memory.

Therefore, in launching a query to the black-box network model, 
the adversary should choose a batch size sufficiently large to activate a
sufficient number of GPU cores to produce a sufficient signal-to-noise ratio. 
We find that the range of the proper batch sizes 
is not prohibitively large (e.g., $64\sim96$ 
for ImageNet networks),
loosely depending on the size of the single input's features and 
the parallel computing ability of the GPU.
In practice, the adversary can choose the batch size by experimenting with
their own GPUs under various image resolutions. 

Notably however, we do not require knowledge of batch size to robustly 
recover network topology (as opposed to hyperparameters), 
only that the batch size is sufficiently large enough to provide a clear signal. 
While we used a consumer friendly sensor with limited 
sampling rate (see \ref{sec:sensorspec}) and corresponding 
signal-to-noise ratio (SNR), a sensor with high sampling rate and 
SNR would correspondingly require a smaller minimum batch size.

\secprespace
\section{Threat Model}\label{sec:threat}
\secpostspace

\paraspace
\paragraph{Incentives.}
An adversary may have numerous motives to carry out a successful model extraction attack.

\textit{Physical IP snooping.}
Most major tech firms provide platforms to manage neural networks as a service
that can be vertically integrated~\cite{nvidiaTriton,abadi2016tensorflow,ibmwatsonpak}.
A client subscribes to the machine learning 
interface
of the service provider but can otherwise
host these services within their own servers to preserve
the integrity of their data.
The ability to host these platforms onsite or on edge devices 
grants local access to 
these
machine learning applications and creates an opportunity for
an adversary who looks to acquire intellectual property. 

\textit{Circumventing payment.}
Neural inference engines
are often deployed as black-box services which an attacker may
have economic incentives to sidestep~\cite{googleaiprice,amazonECSprice,nvidiaTriton}.
In these scenarios, a service provider may invest significant time
and resources towards developing a robust model, 
which it hopes to recover and profit from by charging clients for queries.
Through model extraction and offline training an attacker can avoid
both the development and prediction charges of the reverse engineered model.

\textit{Violating sentries.}
In cases where a machine learning approach is taken to identify
viruses or spam, an attacker increases their chances of deceiving or bypassing
the model by recovering its underlying neural network architecture.
Proximate models have shown to produce better transfer attacks, allowing
one to uncover the classification pitfalls of a similar model
offline before attempting to fool the production model.

For these reasons, enterprises 
that embed networks into their 
applications rely on the privacy of their models.
While the host hardware and operating system may be
secured
by passwords and file access controls, there remains the
threat that an attacker gains physical proximity to the hardware.
{\color{green}
The dangers associated with physical access have been established
by several side-channel works exhibiting different goals under various computing environments~\cite{wei2018know,xiang2020open,weissbart2019one,batina2019csi,dubeymaskednet}.
}
Our study explores how the risk of forfeiting confidentiality
remains present for complex designs running on a GPU.

\paraspace
\paragraph{Target scope.}
Our primary focus centers on using an electromagnetic side channel to 
reverse engineer the neural architecture and its defining layer parameters.
The networks in question may be of arbitrary size and depth, and involve
combinations of fully connected, convolutional, and recurrent layers,
along with a medley of interspersed activation, normalization, and pooling
layers.

Together these layers span the basic components used to assemble most
state of the art networks and account for models used across a variety of 
machine learning 
applications~\cite{he2016vision,xu2017security,teufl2010nlp,kober2013robotics,Simonyan2014vgg,krizhevsky2012imagenet}. 
Furthermore, there are no restrictions on the types of variables involved in the
network, suggesting our method is type agnostic and can support
binary, integer, or real-valued models equally. 

\paraspace
\paragraph{Attacker's capability.}

We follow the threat model of prior works
who measure physical EM signals from models where the adversary may control the
input~\cite{batina2019csi,yu2020deepem} .
Our attacker has no prior knowledge of the target neural network.
The model is taken to be developed, trained, and validated elsewhere.
The only accessible result is
an inference engine whose code, memory, and design 
cannot be accessed without tampering with the service or otherwise alerting the provider.
The attacker is both non-invasive and passive, 
working within standard operating procedure and
treating the network as a black-box while providing inputs of known size to the target.
The adversary can only observe 
the side channel information leaked from the targeted hardware. 
Our attacker does not make any effort to circumvent countermeasures,
given that side channel attacks on neural networks
have only recently been attempted on computing accelerators~\cite{hu2020deepsniffer}.

The side channel information is revealed by placing a magnetic induction sensor 
in close proximity to the GPU's power cable, and launching
a query to produce a measurable signal.
{\color{green}
Our attacker's ability to observe the input and acquire signals
matches the assumptions of several studies that explore side-channel leakage~\cite{weissbart2019one,wei2018know,yu2020deepem,batina2019csi}
}
The attacker is otherwise weak, without ability
to execute code on the host CPU and GPU; and without knowledge 
of the input values and output results of the launched queries, only their size. 
Not only that---they also lack direct access to the GPU hardware, 
beyond the proximity to the power cable.
The adversary only requires access to their own GPU hardware and deep
learning framework (e.g., PyTorch, TensorFlow), matching that of the victim in
order to train offline and carry out the attack.
Although it is possible to extract network information from similar
processors in the family of the target GPU (\secref{gpu-transfer}),
best results are achieved when offline computing is performed on the
same platform as the victim.

\secprespace
\section{Signal Analysis \& Network Reconstruction}\label{sec:method}
\secpostspace

We prepare for the attack by training several \emph{recovery} DNN models. 
After the attacker launches a single batch query (whose input and output 
values are irrelevant), we recover structure from the acquired signal in two stages:
\Li topology and \Lii hyperparameters.
To recover topology, a pretrained DNN model associates a \emph{step} to 
every signal \emph{sample}. 
This per-sample classification partitions the signal into segments corresponding to steps. 
We estimate hyperparameters for each \emph{individual} segment in isolation, using a step-specific pretrained DNN model, and resolve inconsistencies  between consecutive segments using an integer program.
The pretraining of our recovery DNN models is hardware-specific, 
and good recovery requires data gathered from similar hardware. 


\secprespace
\subsection{Topology Recovery}\label{sec:topo}
\secpostspace

%
%

Classifying steps in a network model requires taking in a time-series signal and converting it to labeled operations. 
The EM signal responds only to the GPU's instantaneous performance, 
but because the GPU executes a neural network sequence, there is rich context
in both the window before and after any one segment of the signal.
Some steps are often followed by others, such as pooling operations after a series of convolutions. 
We take advantage of this bidirectional context in our sequence to sequence classification problem by utilizing a recurrent neural network to classify the observed signal. 

Bidirectional Long Short-Term Memory (BiLSTM) networks are well-suited 
for processing time-series signals~\cite{graves2005bidirectional}. 
We train a BiLSTM network to classify each signal sample ${s}_i$ 
predicting the step $C({s}_i)$ that generated ${s}_i$ (see \figref{signalClass}-b). 
The training dataset consists of annotated 
signals constructed automatically (see~\secref{DataConstr}). 
The input signal is first normalized before undergoing processing with a two-layer
BiLSTM network, using a dropout layer of $0.2$ in between.
The input to our network is a sliding window of the time-series signal,
the entirety of which is classified according to the available step operations
from our supervised learning dataset.
For all experiments in our work, we used a layer size of $128$ for the two
BiLSTM layers and an input-output window size of $128$.
We train the BiLSTM by
minimizing the standard cross-entropy loss between the predicted per-sample
labels and the ground-truth labels.
This approach proves robust, and is the method used by all of our
experiments and on all GPU's tested.

The segmented output of our BiLSTM network on our extracted signal is 
for the most part unambiguous. Operations that follow one another 
(i.e. convolution, non-linear activation function, pooling) 
are distinct in their signatures and easily captured from the 
context enabled by the sliding window signal we use as input 
to the BiLSTM classifier. 
Errors that arise come primarily from traces of very small-sized steps, 
closer to our sensor’s sampling limit. 
Noise in such regions may over-segment a non-linear 
activation, causing it to split into two (possibly different) 
activation steps. 
To ensure consistency we post-process the segmented 
results to merge steps of the same type that are output in sequence, 
cull out temporal inconsistencies such as pooling before a non-linear 
activation, and remove activation functions that are larger than 
the convolutions that precede them. 

This concludes identifying the sequence of steps, 
recovering the layers of the 
network, including their \emph{type} 
(e.g., fully connected, convolution, recurrent, etc.), 
activation function, and any subsequent 
forms of pooling or batch normalization. 
What remains is to recover layer hyperparameters. 

\secprespace
\subsection{Hyperparameter Estimation}\label{sec:hyper}
\secpostspace
\paragraph{Hyperparameter consistency.} The number of hyperparameters that 
describe a layer type depends 
on its linear step. For instance, a CNN layer type's linear
step is described by
size, padding, kernel size, number of channels, and stride hyperparameters.
Hyperparameters within a layer must be \emph{intra-consistent}. Of the six CNN 
hyperparameters (stride, padding, dilation, input, output, and kernel size),
any one is determined by the other five.
Hyperparameters must also be \emph{inter-consistent} across consecutive layers:
the output of one layer must fit the input of the next.
A brute-force search of consistent hyperparameters easily becomes
intractable for deeper networks; we therefore first estimate
hyperparameters for each layer in isolation, 
and then jointly optimize to obtain consistency.

\paraspace
\paragraph{Initial estimation.}
We estimate a specific hyperparameter of a specific layer type,
by pretraining a DNN. We pretrain a suite of such DNNs, 
one for each (layer type, hyperparameter) pairing. 
Once the layers (and their types) are recovered, we 
estimate each hyperparameter using these
pretrained (layer type, hyperparameter) recovery DNNs. 

Each DNN is comprised of two 1024-node fully connected layers with dropout.  
The DNN accepts two (concatenated) feature vectors describing two signal
segments: the linear step and immediately subsequent step.
The subsequent step (e.g., activation, pooling, batch
normalization) tends to require effort proportional to the linear step's 
output dimensions, thus its inclusion informs
the estimated output dimension. 
Each segment's feature vector is assembled by
\Li partitioning the segment uniformly into $N$ windows, and computing
the average value of each window,
\Lii concatenating the time duration of the segment.
The concatenated feature vector has a length of $2N+2$.

The DNN is trained with our automatically generated dataset (see \secref{DataConstr}).
The choice of loss function depends on the hyperparameter type:
For a hyperparameter drawn from a wide range, such as 
a \emph{size}, we minimize
mean squared error between the predicted size and the 
ground truth (i.e., regression).
For a hyperparameter drawn from a small discrete distribution,
such as \emph{stride}, we minimize the cross-entropy loss between 
the predicted value and the ground truth (i.e., classification). 
In particular, we used regression for sizes, and classification 
for all other parameters.

\paraspace
\paragraph{Joint optimization.}
The initial estimates of the hyperparameters are generally not \emph{fully} accurate, nor consistent.
To enforce consistency, we jointly optimize all hyperparameters, 
seeking values that
\emph{best fit their initial estimates, subject to consistency constraints}.
Our optimization minimizes the convex quadratic form
\begin{equation}\label{eq:opt}
    \min_{x_i\in\mathbb{Z}^{0+}}\sum_{i\in\mathcal{X}} \left(x_i - x^*_i\right)^2 \ ,\quad
    \textrm{subject to consistency constraints,}
\end{equation}
where $\mathcal{X}$ is the set of all hyperparameters across all layers; $x^*_i$ and $x_i$
are the initial estimate and optimal value of the $i$-th hyperparameter, respectively.
The imposed consistency constraints are:
\begin{enumerate}[topsep=0mm, partopsep=0mm, leftmargin=7mm, label=(\roman*)]
    \item The output size of a layer agrees with the input size of the next next layer.
    \item The input size of the first layer agrees with the input feature size.
    \item The output size of a CNN layer does not exceed its input size (due to convolution).
    \item The hyperparameters of a CNN layer satisfy
          \begin{equation}\label{eq:const}
              s_{\text{out}} = \left\lfloor{\frac{s_\text{in}+2\beta-\gamma(k-1)-1}{\alpha}+1}\right\rfloor,
          \end{equation}
          where $\alpha$, $\beta$, $\gamma$, and $k$ denote the layer's stride, padding, dilation, and kernel size, respectively.
    \item Heuristic constraint: the kernel size must be odd.
\end{enumerate}
Among these constraints, \textbf{(i-iii)} are linear constraints, 
which preserves the convexity of the problem.
The heuristic \Lv can be expressed as a linear constraint: 
for every kernel size parameter $k_j$, we introduce a dummy variable
$\tau_j$, and require $k_j = 2\tau_j+1$ and $\tau_j\in \mathbb{Z}^{0+}$.
Constraint \Liv, however, is troublesome, because the appearance of stride $\alpha$
and dilation $\gamma$, both of which are optimization variables, make the 
constraint nonlinear.

Since all hyperparameters are non-negative integers,
the objective must be optimized via integer programming: 
IP in general case is NP-complete~\cite{papadimitriou1998combinatorial}, 
and the nonlinear constraint \Liv does not make life easier.
Fortunately, both $\alpha$ and $\gamma$ have very narrow ranges in practice:
$\alpha$ is often set to be 1 or 2, and $\gamma$ is usually 1, and they rarely change
across all CNN layers in a network.
As a result, they can be accurately predicted by our DNN models; we therefore retain
the initial estimates and do not optimize for $\alpha$ and $\gamma$,
rendering (\ref{eq:const}) linear.
Even if DNN models could not reliably recover $\alpha$ and $\gamma$, 
one could exhaustively enumerate the few possible $\alpha$ and $\gamma$ combinations,
and solve the IP problem (\ref{eq:opt}) for each combination, and select the best recovery.

The IP problem with a quadratic objective function and linear constraints can be easily
solved, even when the number of hyperparameters is large (e.g., $>1,000$). 
In practice, we use IBM CPLEX~\cite{cplex2009v12}, a widely used IP solver.
Optimized hyperparameters remain close to the initial DNN estimates,
and are guaranteed to define a valid network structure.

\secprespace
\section{Experimental Setup}\label{sec:setup}
\secprespace

In the following section we discuss hardware choices, sensor setup, and dataset
generation details of our experiments.

\begin{figure*}[ht]
    \centering
    \vspace{-1mm}
    \includegraphics[width=0.98\textwidth]{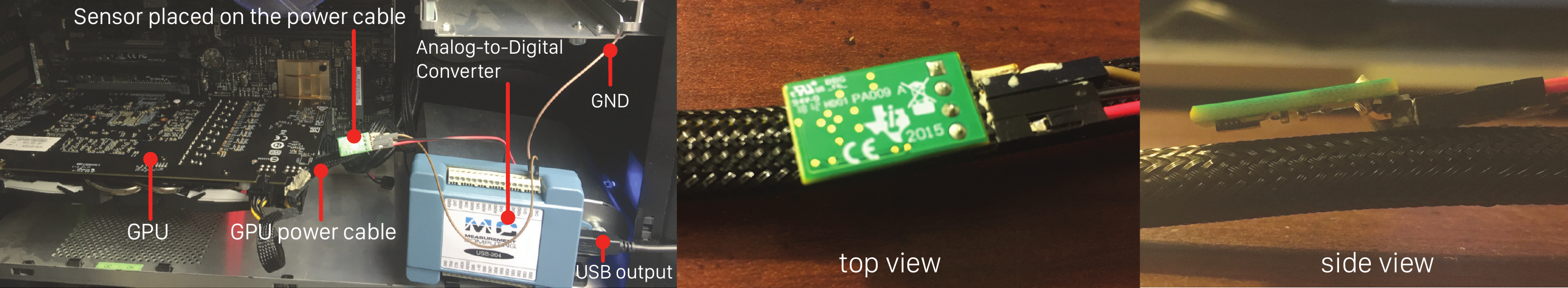}
    \vspace{-2mm}
    \caption{ \textbf{Sensing setup.} Placement of the magnetic induction sensor on the power cord works regardless of the GPU model, providing a common weak-spot to enable current-based magnetic side-channel attacks.}
    \label{fig:physicalSensor}
    \vspace{-2mm}
\end{figure*}

\subsection{Hardware Sensors}\label{sec:sensorspec}
\secpostspace

We use the DRV425 fluxgate magnetic sensor from Texas Instruments for
reliable high-frequency sensing of magnetic signals~\cite{drv425,petrucha2018testing}.
This sensor, though costing only \$3 USD, outputs robust analog 
signals with a 47kHz sampling rate and $\pm$2mT sensing range.
For analog to digital conversion (ADC), 
we use the USB-204 Digital Acquisition card, 
a 5-Volt ADC from Measurement Computing~\cite{adc}.  
This allows a 12-bit conversion of the signal, 
mapping sensor readings from -2mT$\sim$2mT to 0V$\sim$5V.  

{\color{green}
Previous works have attempted electromagnetic side-channel attacks with industrial probes, 
e.g. the Langer RF-U 5-2~\cite{batina2019csi}. 
These sensors are rated to measure at higher frequencies ranging from 30MHz to 3GHz, 
but are far more expensive (\$1,500+ USD) and require additional technical equipment to operate. 
Other power side channel exploits have explored sampling at again higher rates 
varying from 400KHz to 2.5GHz~\cite{wei2018know,xiang2020open}. 
Sampling at these frequencies allows finer capture of hardware signals 
that can ease network step classification, 
although it significantly increases the amount of feature data to be processed. 
In contrast, our method involves a \$3 USD sensor sampling at 47KHz. 
Sampling at rates lower than 47KHz would make it difficult 
to adequately capture short events on the GPU,
such as non-linear activation functions and small matrix multiplications that may occur.
}
\secprespace
\subsection{Dataset Construction}\label{sec:DataConstr}
\secpostspace

\paragraph{Sensor placement.}
Setup of the sensor requires that \Li the 
sensor is within range of the electromagnetic signal and \Lii the 
sensor orientation is consistent.
To avoid interference from other electric components,
we place the sensor near the GPU's magnetic induction 
source, anywhere along the power cable. 
Because magnetic flux decays inversely proportional 
to the squared distance from the source, 
according to the Biot-Savart law~\cite{griffiths2005introduction},
we position the sensor within millimeters of the cable casing (see \figref{physicalSensor}). 
Flipping the sensor over will result in a sign change of the received magnetic 
induction signal, thus we maintain a uniform orientation to avoid the misalignment of readings across the dataset.

\paraspace
\paragraph{Data capture.}
Pretraining the recovery DNN models (recall \secref{method})
requires an annotated dataset with pairwise correspondence between signal and 
step types (see \figref{physicalSensor}).
We can automatically generate an annotated signal for a given network
and specific GPU hardware, simply by executing a query (with arbitrary input values) on the 
GPU to acquire the signal. 
Timestamped ground-truth GPU operations
are made available by most deep learning libraries (e.g., \texttt{\small torch.autograd.profiler} in PyTorch and
\texttt{\small  tf.profiler} in TensorFlow). 

A difficulty in this process lies in the fact that
the captured (47kHz) raw signals and the ground truth GPU traces run on
different clocks.
Similar to the use of clapperboard to synchronize picture and sound in filmmaking, 
we precede the inference query with a short intensive GPU operation to
induce a sharp spike in the signal,
yielding a synchronization landmark (see \figref{delta}). 
We implemented this ``clapperboard'' by filling a
vector with random floating point numbers.

\paraspace
\paraspace
\paragraph{Training Set Details.} 
The set of networks to be annotated could in principle consist \Li solely of 
randomly generated networks, on the basis that data values and ``functionality'' are irrelevant 
to us, and the training serves to recover the substeps of a layer;
or \Lii of curated networks or those found in the wild, on the basis that such 
networks are more indicative of what lies within the black-box.

We construct our training set as a mixture of both approaches.
Randomly generated networks involve base steps made up of a mixture of fully-connected, recurrent, and CNN layers.
These are accompanied by $5$ different activation functions, $2$ types of pooling layers,
and a potential normalization operation.
Off the shelf networks consist of VGG and ResNet variants.
All in all we consider $500$ networks for training,
ranging from $4$ to $512$ steps per network and culminating in
$70,933$ individual steps in total.
When we construct these networks, their input image resolutions are randomly chosen from
[224$\times$224, 96$\times$96, 64$\times$64, 48$\times$48, 32$\times$32]: the highest resolution
is used in ImageNet, and lower resolutions are used in datasets such as CIFAR.
We will release training and test datasets along with source code and hardware schematics for full reproducibility.

\paraspace
\paragraph{Test dataset.}
We construct a test dataset fully separate from the training dataset.
Our test dataset consists of $64$ randomly generated networks produced the 
same way as those randomly generated for training. 
The number of layers ranges from 30 to 50 layers. 
To diversify our zoology of test models, we also include 
smaller networks that are under 10 layers, LSTM networks, as well as 
ResNets (18, 34, 50, and 101).
Altogether, each test network has up to 514 steps. In total, the test dataset includes $5,708$ 
network steps,
broken down into 
$1,808$ activation functions, 
$1,975$ additional batch normalization and pooling, and
$1,925$ fully connected, convolutional, and recurrent layers. 


\secprespace
\vspace{-1mm}
\section{Results}\label{sec:ret}
\vspace{-1mm}
\secpostspace

This section presents the major empirical evaluations of our method. 
We refer the reader to \appref{exp} for additional results,  experiments, and discussion.

\secprespace
\vspace{-1mm}
\subsection{Accuracy of Network Reconstruction}
\label{sec:class-accuracy}
\secpostspace

\begin{table}[t]
\vspace{-1mm}
\centering
\caption{Classification accuracy of network steps (Titan V) \label{tab:classify1}}
\vspace{1mm}
\scalebox{0.99}{
 \begin{tabular}{l|cccc} 
\bottomrule
 Layer Type & Prec. & Rec. & F1  & \# samples \\ \hline
LSTM & .997 & .992 & .995 & 8,704 \\
Conv & .993 & .996 & .994 & 447,968  \\
Fully-connected & .901 & .796 & .846 & 10,783 \\
Add & .984 & .994 & .989 & 22,714   \\
BatchNorm & .953 &	.955 &	.954 &	47,440\\
MaxPool & .957 & .697 & .806 & 4,045 \\
AvgPool & .371 & .760 & .499 &675 \\
ReLU  & .861 & .967& .911 & 28,512\\
ELU  & .464 &.825 & .594& 2,834 \\
LeakyReLU&.732 & .578 & .646 & 9,410 \\
Sigmoid& .694 & .511 & .588& 8,744 \\
Tanh& .773& .557& .648& 4,832 \\ \hline
Weighted Avg. & \textbf{.968}& \textbf{.967} & \textbf{.966} & - \\
\toprule
 \end{tabular}}
\vspace{-3mm}
\end{table}

\paragraph{Topology reconstruction.}
As discussed in~\secref{method}, we use a BiLSTM model
to predict the network step for each single sample.
\tabref{classify1} reports its accuracy, measured on an Nvidia Titan V GPU. 
There, we also break the accuracy down into measures of individual types of network steps, with an overall accuracy of \textbf{96.8\%}.
An interesting observation is that the training and test datasets are both
unbalanced in terms of signal samples (see last column of \tabref{classify1}).
This is because in practice convolutional layers are computationally the most
expensive, while activation functions and pooling are lightweight.
Also, certain steps like average pooling are much less frequently used.
While such data imbalance does reflect reality, when
we use them to train and test, most of the misclassifications occur at those
rarely used, lightweight network steps, whereas the majority of network steps
are classified correctly.

We evaluate the quality of topology reconstruction using
normalized Levenshtein distance (i.e., one of the edit distance metrics) that 
has been used to evaluate network structure similarity \cite{graves2006connectionist, hu2020deepsniffer}.
Here, Levenshtein distance measures the minimum number of operations---including
adding/removing network steps and altering step type---needed to 
fully rectify a recovered topology. 
This distance is then normalized by the total number of steps of the target network.

We report the detailed results in Figure~\ref{fig:distancetitanv} in the appendix. Among the 64 tested networks, 40 of the reconstructed
networks match precisely their targets, resulting in \emph{zero} Levenshtein distance.
The average normalized Levenshtein distance of all tested networks is \textbf{0.118}.
This confirms our networks are recovered with similar network lengths and often exact
step matches.

\begin{figure}[t]
  \begin{center}
    \includegraphics[width=0.47\textwidth]{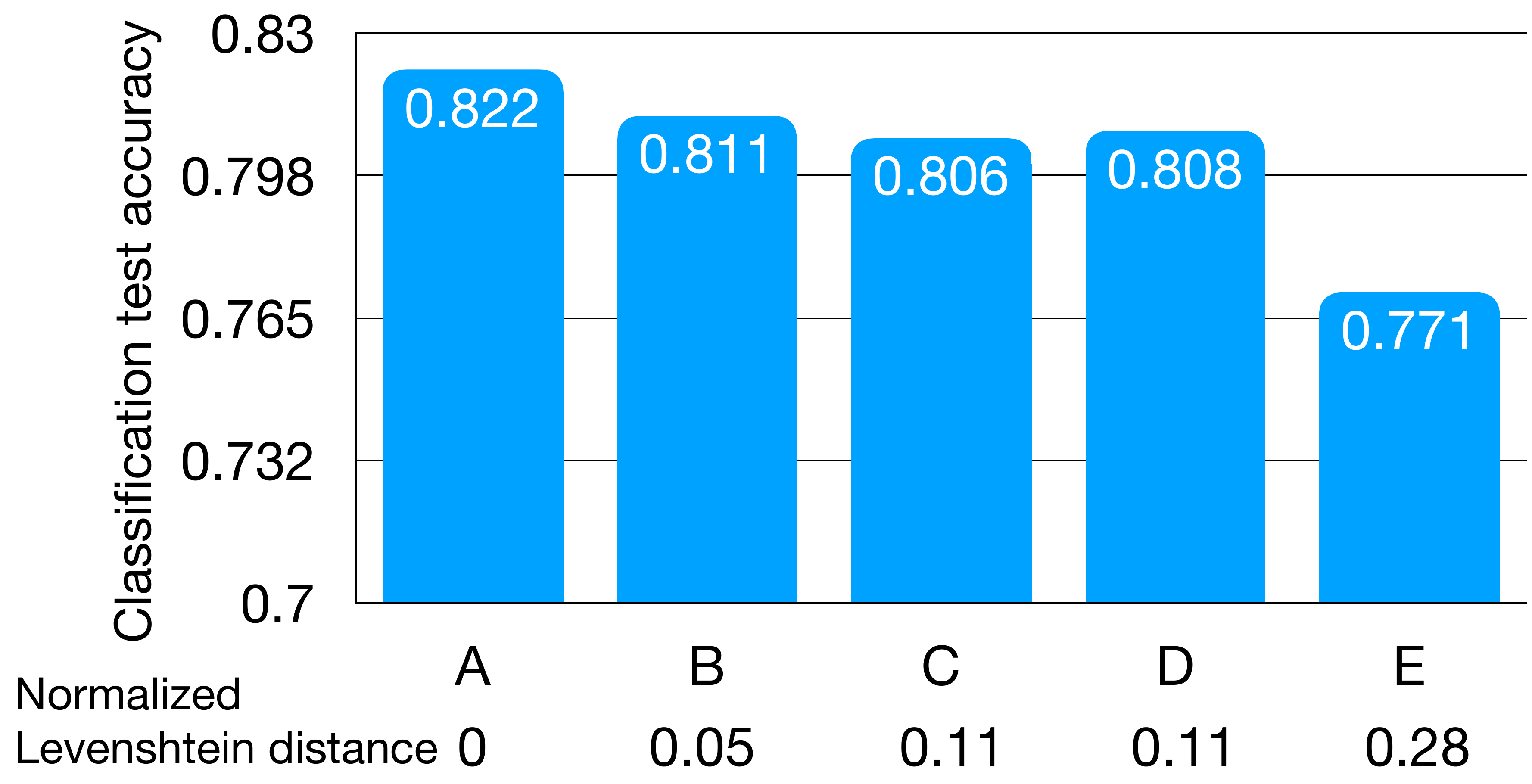}
  \end{center}
  \vspace{-6mm}
  \caption{Each model's classification accuracy drops as its Levenshtein distance
      from the original model (model \textsf{A}: AlexNet) increases. \label{fig:edit_dist}}
  \vspace{-2mm}
\end{figure}

To provide a sense of how the normalized Levenshtein distance is related to a network's ultimate performance, we
conduct an additional experiment to gauge reconstruction quality via classification accuracy.
We consider AlexNet (referred as model \textsf{A}) and its
five variants (refered as model \textsf{B}, \textsf{C}, \textsf{D}, and \textsf{E}, respectively).
The variants are constructed by randomly altering some of the network steps in model \textsf{A}.
The Levenshtein distances between model \textsf{A} and its variants are 1, 2, 2, 5, 
respectively, and the normalized Levenshtein distances 
are 0.05, 0.11, 0.11, 0.28 (see \figref{edit_dist}). 
We then measure the performance (i.e., standard test accuracy) of these models
on CIFAR-10. As the edit distance increases, the model's performance drops.

\paraspace
\paragraph{DNN hyperparameter estimation.}
Next, we report the test accuracies of our DNN models (discussed in
\secref{hyper}) for estimating hyperparameters of convolutional layers. 
Our test data here consists of 1804 convolutional layers. On average, our DNN
models have \textbf{96\%-97\%} accuracy.  The break-down accuracies for
individual hyperparameters are shown in \tabref{dnn} of the appendix.

\begin{table}[t]
\centering
\caption{Model extraction accuracy on CIFAR-10 
}\label{tab:cifarAcc}
\resizebox{\columnwidth}{!}{%
 \begin{tabular}{l|ccccc}
\bottomrule
{Model}
 & Target & Titan V & Titan X & GTX1080 & GTX960 \\ \hline
 VGG-11 & 89.03 & 89.61  & 89.63 & 88.46 & 88.3\\
 VGG-16 & 90.95 & 91.08 & 91.03 & 89.33 & 90.78\\
 AlexNet & 81.68 & 85.26 & 85.11 & 85.27 & 85.03\\
 ResNet-18 & 92.77 & 92.61 & 92.82 & 92.79 & 92.04\\
 ResNet-34 & 92.21 & 92.28 & 92.95 & 90.81 & 92.71\\
 ResNet-50 & 90.89 & 91.8 & 91.97 & 91.2 & 91.29 \\ 
 ResNet-101 & 91.58 & 91.91 & 91.85 & 91.37 & 91.72 \\ 
\toprule
 \end{tabular}
 }
 \vspace{-4mm}
\end{table}

\paraspace
\paragraph{Reconstruction quality measured as classification accuracy.}
Ultimately, the reconstruction quality must be evaluated
by how well the reconstructed network performs in the task that the 
original network aims for. To this end, we test seven networks, including VGGs,
AlexNet, and ResNets, that have been used for CIFAR-10 classification (shown in
Table~\ref{tab:cifarAcc}). 
We treat those networks as black-box models and reconstruct them from their magnetic signals. 
We then train those reconstructed networks and compare their test accuracies with the original networks' performance.
Both the reconstructed and original networks are trained with the same training dataset for the same number of epochs. 
The results in Table~\ref{tab:cifarAcc} show that for all seven networks, including large networks (e.g., ResNet101), the reconstructed networks perform almost as well as their original versions. 
We also conduct similar experiments on ImageNet and report the results in \tabref{imagenet} of \appref{imagenet}.

\secprespace
\subsection{Accuracy across GPUs}
\label{sec:gpu-transfer}
\secpostspace

\paragraph{Twin GPU transferability.}

\begin{table}[t]
\vspace{-1mm}
\centering
\caption{Classification accuracy of network steps (GTX-1080).
    \label{tab:valid-gtx}}
\vspace{-2mm}
\resizebox{\columnwidth}{!}{
 \begin{tabular}{l|cccc} 
\bottomrule
 & Prec. & Rec. & F1  & \# samples \\ \hline
LSTM&  .997 & .999 & .998 & 12,186 \\
Conv & .985 & .989 & .987 & 141,164 \\
Fully-connected &  .818 & .969 & .887 & 9,301 \\
Add & .962 & .941 & .951 & 30,214 \\
BatchNorm & .956 & .944 & .950 & 48,433\\
MaxPool  & .809 & .701 & .751 & 1,190\\
AvgPool  & .927 & .874 & .900 & 294\\
ReLU   & .868 & .859 & .863 & 11,425\\
ELU   & .861 & .945 & .901 & 8,311\\
LeakyReLU & .962 & .801 & .874 & 3,338\\
Sigmoid & .462 & .801 & .585 & 5,106\\
Tanh&  .928&	.384& .543&	8,050\\ \hline
Weigted Avg.  & \textbf{.945}	&\textbf{.945}&\textbf{.945}& -\\
\toprule
 \end{tabular}
 }
 \vspace{-1mm}
\end{table}

\begin{figure*}[t]
    \centering
    \vspace{-3mm}
    \includegraphics[width=0.9\textwidth]{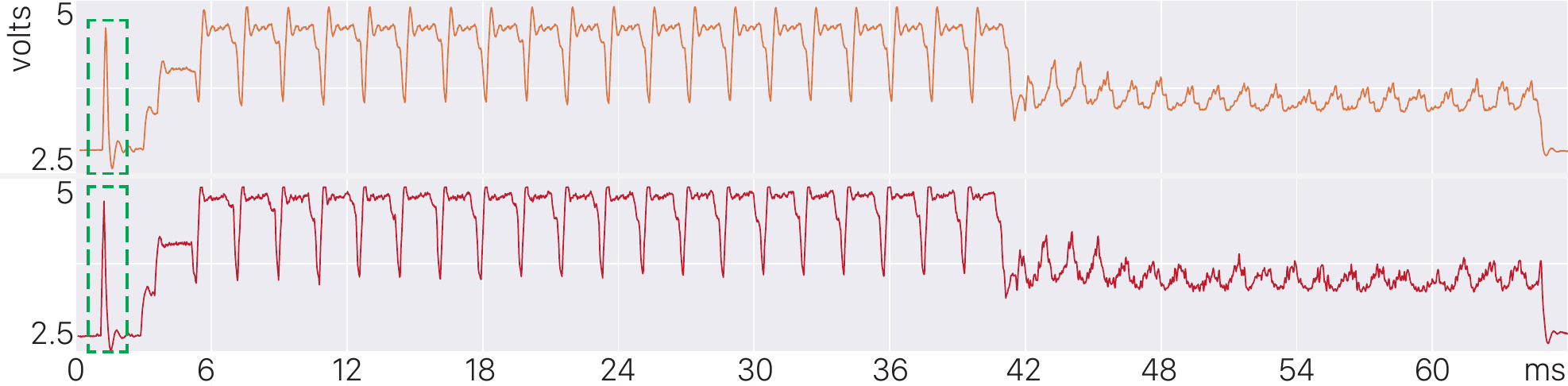}
    \vspace{-3mm}
    \caption{Here we plot the resulting signals from the same network model deployed on
    two different instances of a Nvidia GTX 1080 (running on two different computers). In 
    the green dash boxes on the left are the spikes that we inject on purpose (discussed in \secref{DataConstr}) 
    to synchronize the measured signal with the runtime trace of the GPU operations.}
    \label{fig:delta}
    \vspace{-2mm}
\end{figure*}

Our proposed attack requires the adversary to have
the same brand/version of GPU as the victim, but not necessarily the same physical copy (see \figref{delta}). 
Here, we obtain two Nvidia GTX-1080 Graphics cards running on two different machines with different CPUs and RAM, using one to generate training data and another one for black-box reconstruction. 

We set out to verify that \Li the leaked magnetic signals 
are largely related to GPU brand/version but not the other factors such as CPUs and \Lii the signal characteristics from two physical copies of the same GPU type 
stay consistent. 
When we run the same network structure on both GPUs, the resulting magnetic signals are 
similar to each other, as shown in Figure~\ref{fig:delta}.
This suggests that the GPU cards are indeed the primary sources contributing the captured magnetic signals.

Next, we use one GPU to generate training data and another one to collect signals and test our
black-box reconstruction. 
The topology reconstruction results are shown in  \tabref{valid-gtx}, arranged in the way similar to \tabref{classify1}, and 
the distribution of normalized Levenshtein edit distance over the tested networks are shown 
in Figure~\ref{fig:distancegpu} of the appendix.
These accuracies are very close to the case wherein a single GPU is used.
The later part of the reconstruction pipeline (i.e., the hyperparameter recovery) directly depends on the topology reconstruction. 
Therefore, we expect that the final reconstruction is also very similar to the single-GPU results.

\paraspace
\paragraph{Unspecified GPU transferability.}
Thus far we have considered the case where the attacker knows the hardware model and version of the target GPU.
We now explore the case where the model and version are unavailable. 
The adversary may attempt to address this by exhaustively 
pretraining reconstruction networks for a wide range of GPU models. 
The signal collected from the target GPU may then be processed 
by the wide range of reconstruction networks. 
Since software upgrades may affect GPU performance  
and alter the emitted magnetic signals, 
the software version is considered to be part of the GPU model specification.
In our experiments, we keep software versions constant, 
including OS version, CUDA version, and PyTorch/Tensorflow version.

The scenario in which GPU model and version are unavailable 
can present itself in two ways:
The victim GPU may or may not be represented during the network training phase.
We explore each case in turn.

Suppose that the victim GPU is not represented in the network training phase.
To explore this scenario, we perform K-fold 
cross-validation of classification accuracy across $6$ unique GPUs.
For each potential victim GPU, 
we perform reconstruction multiple times, each time against measurements of another GPU, utilizing a new dataset comprised of all other GPUs. 
The results of this K-fold cross-validation can be found on the top row of \tabref{unspecifiedGPUs}.
There we find that older GPUs (such as the GTX-960) perform poorly when 
trained on newer counterparts.
Accuracy improves when trained with similar models, such as the three 1080 architectures,
which although they have different memory allowances and hardware specifications,
result in high holdout accuracy.
This suggests that it is possible to achieve favorable results even without precise knowledge
of the target GPU, so long as the training employs a similar GPU.

Next, suppose that the victim GPU is represented in the network training phase.
Although our proposed attack encourages the attacker to train solely and specifically
with the victim GPU, by assumption this is not possible in this scenario. 
Instead, we train one generalized network comprised of signals from all available GPUs.
We then evaluate the reconstruction accuracy for each specific GPU 
against the generalized network. As recorded in the second row of~\tabref{unspecifiedGPUs},
reconstruction accuracy improves when the victim GPU is represented during training, even if the training set is ``contaminated'' by data from other GPU models.

\paraspace
\paragraph{Multi-GPU Workstations.}
Multi-GPU configurations do not introduce interference so long as their 
power cables are isolated and sufficiently far apart (>7mm). 
An easy way to avoid any such interference is to 
record the signal either on the GPU itself or proximate to the connection between
power cable and GPU. 
We experimented with recording multi-GPU configurations
and found no issues isolating the side-channel information of the victim GPU.


\vspace{-2mm}
\begin{table*}[t]
\centering
\caption{Classification accuracy on datasets combining many GPUs}
\label{tab:unspecifiedGPUs}
 \begin{tabular}{l|cccccc} 
\bottomrule
& \multicolumn{6}{c}{Target GPU} \\
& GTX-960 & MSI-1060 & MSI-1070 & MSI-1080 & EVGA-1080 & GTX-1080 \\ \hline
With Holdout & 61.3 & 77.4 & 83.4 & 87.1 & 93.2 & 93.9 \\ 
Full Dataset & 96.5 & 88.6 & 93.4 & 91.7 & 95.8 & 95.2 \\ 
\toprule
 \end{tabular}
\end{table*}

\secprespace
\subsection{Transfer Attack}
\label{sec:transferability}
\secpostspace

An adversarial transfer attack attempts to design an input that tricks an unknown target model. The name 
\emph{transfer} alludes to the method of attack: The attacker builds a surrogate, an approximation (informed guess) of the unknown target model, and seeks out an input that tricks the surrogate. The attacker hopes that the exploit ``transfers'' to the actual target, i.e., that an input that tricks the surrogate also tricks the target. 
The likelihood of a successful attack increases as the surrogate better approximates the target. In a black-box setting, finding an effective surrogate is very hard~\cite{demontis2019adversarial}. 
Therefore, the attacker wishes to design a more informed surrogate. One avenue toward this is to design surrogates with topology and parameters similar to the target.

\paraspace
\paragraph{CIFAR-10 Dataset.}
Here we test on six networks found in the wild, 
ranging from VGGs to AlexNet to ResNets, as listed on the header row of
\tabref{cifar}. The table shows the percent of successful transfer 
attacks over $5,000$ attempts on the CIFAR-10 dataset.

We consider each target architecture on each of four GPUs in turn (top four rows of \tabref{cifar}). 
We consider each such architecture-GPU combination, in turn, as black-box target. 
Using the side channel exploit, we reconstruct the target's structure to 
obtain a surrogate architecture, 
which we train on CIFAR-10 to obtain a surrogate model. 
We craft inputs that trick the surrogate, and evaluate whether those inputs also trick the target. 
Transfer attack success is defined as the percent of generated inputs (based on the surrogate) that correctly cause the trained target model to mislabel an input.
All adversarial inputs are generated via Projected Gradient Descent~\cite{madry2018towards}, using an $\epsilon$ of $0.031$ and an $\alpha$ of $0.003$ for all results.
The success rate of the transfer attacks is summarized in the upper four rows of \tabref{cifar}. 


To gauge the success rates of the ``side channel surrogates,'' we compare them against ``white-box surrogates.'' We build six white-box surrogates, corresponding to the six known target architectures; these white-box surrogates differ only in weights, as the surrogates are trained from scratch on CIFAR-10.
The idealized white-box surrogates serve as a benchmark for effective surrogates; refer to the success rates in the bottom six rows of \tabref{cifar}.

Remarkably, the ``side-channel surrogates'' offer comparable
success rates to
``white-box surrogates.'' 
The relative success of side-channel surrogates becomes more pronounced for deeper networks (ResNets), where it appears that architecture dominates sensitivity to weight values.
When the number of layers is small, as in VGG-11 and AlexNet architectures, 
the margin for error decreases, 
and more importance is given to the weights of the target. 
However, even in these cases where attack performance drops, 
the side-channel surrogates closely match the success rate 
of their white-box counterparts, 
displayed in the lower rows. 
In other words, the side-channel reconstruction effectively 
turns a black-box into a white-box attack.

\begin{table*}[t]
\centering
\vspace{-1mm}
\caption{Transfer attack results on CIFAR-10.\label{tab:cifar}}
\begin{tabular}{cl|cccccc}
\bottomrule
 & & \multicolumn{6}{c}{Target Model} \\
& & ResNet-18 & ResNet-34 & ResNet-101 & VGG-11 & VGG-16 & AlexNet \\\hline
\parbox[t]{2mm}{\multirow{11}{*}{\rotatebox[origin=c]{90}{Source Model}}} 
& GTX-960&  98.56 & 92.51 & 91.20 & 63.41 & 72.57 & 58.90\\
& GTX-1080& 97.88 & 90.86 & 86.24 & 64.69 & 55.19 & 56.83 \\
& Titan X& 98.32 & 93.45& 84.47 & 61.89 & 77.36 & 68.41\\
& Titan V& 98.48 & 93.65 & 91.27 & 64.39 & 72.77 & 60.17\\\cline{2-8}
& ResNet-18& 97.70  & 90.72 & 80.27 & 47.98 & 86.64 & 30.56\\
& ResNet-34 & 97.21 & 92.46 & 82.30 & 51.42 & 85.60 & 32.34\\
& ResNet-101 & 92.53 & 86.98 & 92.95  & 53.98 & 83.04 & 30.55\\
& VGG-11& 65.86 & 57.82 & 57.52 & 60.24 & 65.50 & 39.95\\
& VGG-16& 74.00 & 61.54 & 54.23 & 41.60 & 74.29 & 29.57\\
& AlexNet & 10.11 & 9.59& 10.19 & 11.60 & 10.42 &  62.70\\
\toprule
\end{tabular}
\vspace{-2mm}
\end{table*}

\paraspace
\paragraph{MNIST Dataset.}
Similar to our analysis of CIFAR-10 transfer attacks, 
we also conduct transfer attack experiments on the MNIST dataset. 
We download four networks online, which are not commonly used.
Two of them are convolutional networks (referred as CNN1 and CNN2), and 
the other two are fully connected networks (referred as DNN1 and DNN2).
None of these networks appear in the training dataset.
We treat these networks as black-box targets, reconstruct a side-channel surrogate for each, and attack the four targets; results are shown in \tabref{mnist}.
As baselines, we also train white-box surrogates with the exact architecture of the four target models.

All four of our extracted networks, visible in the top row of \tabref{mnist} achieve high
transfer attack scores against our candidate targets.
These high scores suggest a close approximation of the target models by our reconstructed networks.
The similarity between our extracted network's transfer attack results and the results achieved
by the matching source model across the bottom four rows also indicates a strong correspondence
in the achieved architectures.
We find that even across the MNIST dataset we are able to generate a model that behaves akin to
a white-box transfer attack.

\vspace{-2mm}
\begin{table}[t]
\centering
\caption{Transfer attack results on MNIST. }\label{tab:mnist}
\resizebox{\columnwidth}{!}{
 \begin{tabular}{ll|cccc} 
\bottomrule
& & \multicolumn{4}{c}{Target Model} \\
& & CNN1 & CNN2 & DNN1 & DNN2 \\ \hline
\parbox[t]{2mm}{\multirow{5}{*}{\rotatebox[origin=c]{90}{Source Model}}} 
& GTX-1080 & .802 & .878 & .999 & .874 \\ 
\cline{2-6}
& CNN1 & .858 & .226 & .785 & .476 \\
& CNN2 & .395 & .884 & .354 & .354 \\
& DNN1 & .768 & .239 & .999 & .803 \\
& DNN2 & .703 & .219 & .975 & .860 \\ 
\toprule
 \end{tabular}
 }
\end{table}

\secprespace
\section{Defenses Against Magnetic Side Channels} \label{sec:defense}
\secpostspace

At this point, we have shown the robustness and accuracy of the magnetic side
channel exploits and turn our attention to countermeasures.  
Traditionally side channel defenses fall under the category of
either detection or reduction of the relevant signal ~\cite{spreitzer2018classSideChannel}.
Since our approach is non-invasive and passive in
that it does not alter any code or hardware operation of GPUs, detection
methods which consist of somehow discovering someone is listening to the
GPU are not applicable to magnetic leakage.
Instead we focus on suppression techniques of the correlated signal,
which aim to decrease the leaked signal-to-noise ratio by either 
confounding the signal in place or concurrently injecting noise to mask emissions~\cite{dasCounter2020}.
We explore both these avenues separately by looking at \emph{prevention} and \emph{jamming}. 

\secprespace
\subsection{Prevention}\label{sec:prevention}
\secpostspace

\begin{figure}[t]
    \centering
    \includegraphics[width=0.99\linewidth]{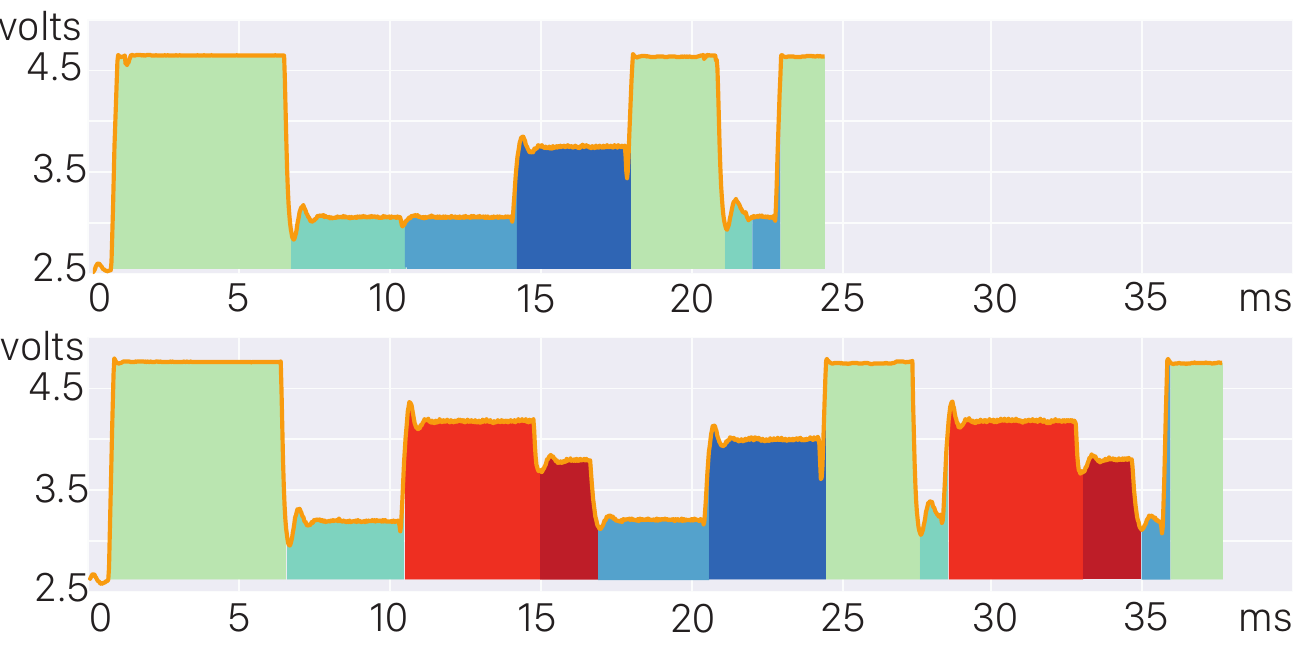}
    \vspace{-2mm}
    \caption{
    \textbf{Deceptive steps.}
    Our side channel cannot track dataflow across the network to distinguish relevant operations (green and blue highlights).
    Extraneous interspersed steps (red and orange) can mix in signals
    to impede topology extraction, trading off less efficient processing for security.
    \label{fig:deadBranch}}
    \vspace{-2mm}
\end{figure}

As shown in Figure~\ref{fig:signalClass}, each rise and drop of the magnetic signals correspond to the boundary between GPU operations. 
This is only possible when the input batch is large enough to keep every GPU operation sustained and stabilized at a high-load state. 
To prevent this behavior, one can restrict the input to be sufficiently small (e.g. 1
single image) whenever appropriate, such that the magnetic signals 
never reach any stable state and 
suffer from a low signal-to-noise ratio,
rendering our sensing setup futile.  

Another way to
prevent magnetic side channel leakage is to use a non-standard framework for
inference which the adversary does not have any training data to start with.
Whether by bringing into play new low-level GPU kernels or operating in unique sequences,
an atypical software implementation of a network architecture will result in magnetic signatures
that are unaccounted for in the offline training. 

Yet another possible defense mechanism results from the fact that we are not
tracking the actual dataflow in the GPU.
For example, we can correctly
identify two GPU operations, convolution and batch norm, within a long sequence.
But there is no evidence to be found within the magnetic side channel proving the dataflow follows the same pattern---the
output from convolution could be a dead end and batch norm takes input from a
previous GPU operation.  
This mismatch between the dataflow and the underlying
network model makes it hard to decipher network measurements robustly.  

We explore this approach in Figure~\ref{fig:deadBranch}, and find that one can
muddle the signal by periodically altering the flow of logic in the target network.
By introducing additional operations \emph{within} the network logic for the GPU
to perform, the network appears to consist of steps that are not in fact
necessary for inference.
We achieve this by regularly performing an additional convolution,
normalization, or pooling function on a copy of the dataflow input at any junction along
the network.
These accessory network steps create dead branches in the topology of the network,
producing data that is abandoned since it is not used anywhere further along as part of the inference.
In our experiments we found that although our BiLSTM had no issues
classifying any invalid steps added, 
they were nevertheless sufficient to derail our estimate of network topology.
However we also confirmed that shuffling in additional computation 
has the undesirable effect of prolonging
inference times and reducing overall network efficiency~\cite{dasCounter2020},
leading to an inference delay linear in the quantity, types, and sizes of the
extraneous steps introduced.


\secprespace
\subsection{Jamming}\label{sec:masking}
\secpostspace

While running on tiny input batch size or tampering with network logic may at times be infeasible, 
we find jamming to be an additional effective defense mechanism when applicable. 
Unlike simpler processors, GPUs are capable of concurrently operating multiple programming kernels (e.g., each in a different CUDA stream).
This allows for alternative code to run adjacent to the hidden target network, 
inducing EM signals that potentially obfuscate the signal related to the model architecture. 

Specifically, during the inference of a large input batch, 
we ran a third-party CUDA GPU stress test in the background~\cite{gpuburn}.
We found that the magnetic signals are completely distorted because 
of the constant high utilization of GPU.
The main caveats opposing this heavy handed approach involve higher power consumption 
and the possible effects on the lifetime of a GPU.


\begin{figure}[t]
    \centering
    \includegraphics[width=0.94\linewidth]{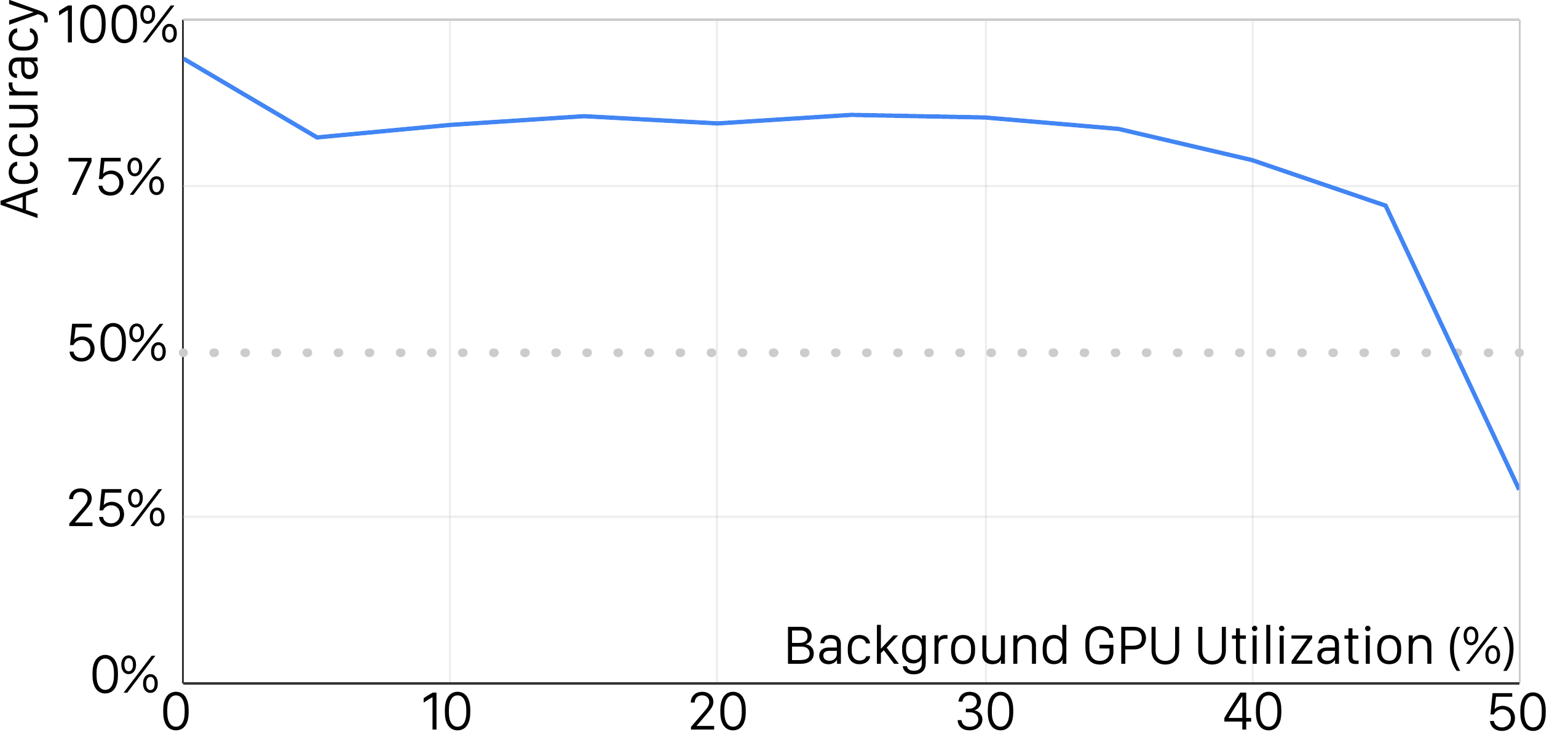}    
    \caption{ 
    \textbf{Classification accuracy in the precense of noise.} We simultaneously run additional immaterial kernels on the GPU while collecting inferences on a target dataset. The background load is increased until accuracy on the target drops below $50\%$.
    \label{fig:maskplot}}
\end{figure}

To further quantify how secondary usage might mask the signal associated to primary usage,
we also trained a network to recover signals from a GTX-1080 GPU
while incrementally adding a background load throughout testing.
The network is solely trained on a dataset of signals free of noise or any 
secondary processes on the GPU. 
Next, we record a fixed test set of signals, repeating the recording process at varying levels of background load on the GPU. We repeat this process until the extraction accuracy falls below $50\%$. The results are depicted in Figure~\ref{fig:maskplot}.

Though ample background use can limit our sensor's effectiveness,
it is important to note that normal background, variable,
and low-utilization GPU operations do not affect our signal recovery.
Figure~\ref{fig:maskplot} shows that the network performs best when free of any noise,
and degrades as concurrent background noise introduces an additional $35\%$ GPU utilization.
Around and beyond this level of noise we find that the background signals
reached similar peaks to that of the target network,
whose unhindered GPU utilization ranges from $10\%\sim45\%$.
Consequently, as the signal becomes dominated by background signals, the accuracy drops, 
until a masking effect akin to jamming with a constant signal
is achieved between $45\%\sim50\%$ background utilization. 

\secprespace
\section{Ethical Considerations}\label{sec:ethics}
\secpostspace






The general notion that a magnetic side channel can leak 
information has been disclosed in prior work~\cite{wei2018know,xiang2020open,weissbart2019one,batina2019csi,dubeymaskednet}.
However, the extent to which such a side channel can recover neural architecture details
is becoming more evident.
As with all disclosures of vulnerabilities, sharing these findings creates the potential for malicious use.
Due to the non-intrusive nature of this attack,
network application logs do not include the requisite details to 
determine via audits whether such malicious use is already occurring. 

At the same time, 
such side channel information 
can lead to positive outcomes, such as
non-intrusive hardware monitoring and intellectual property protection. 

Therefore, it is not straightforward to weigh the costs and benefits of 
disclosing this vulnerability. 
What is clear is that creating a better shared understanding of 
such vulnerabilities is a necessary step toward developing
appropriate precautions, safeguards, and countermeasures.

We have corresponded with vendors to disclose and describe our findings,
and will provide the requisite time for mitigation.
We have also discussed several viable defense mechanisms
targeting GPU processes and network implementations 
in response to our study.

\secprespace
\section{Discussion}\label{sec:conclusion}
\secpostspace
\paragraph{Comparisons.}
Our method considers more general recovery of more complex GPU-based networks using lower cost sensors. 
Compared to prior works~\cite{batina2019csi,yu2020deepem}, 
we recover networks from a single EM trace using a data-driven approach leveraging BiLSTMs, 
rather than requiring multiple scans (potentially tens-of-thousands) 
and relying on additional steps like Correlation Power Analysis or visual inspection for topology recovery. 
Approaches based on these methods make additional assumptions about layer sizes 
and handle a maximum of $7$ and $23$ layers, respectively, 
whereas we accommodate hundreds of layers.
It is unclear whether these methods extend beyond the straightforward and small 
multi-layer perceptrons and CNNs programmable on simpler processors to other layer types and models.
Moreover these prior methods use setups that cost thousands of dollars 
and involve either tampering with the chips or carefully positioning processors against sensors via stepper-motors, 
particularly limiting application to GPUs.

Unlike previous approaches, 
we do not require access to input-output variables 
and can handle large state-of-the-art models using a \$$3$ sensor. 
Our side-channel exploit stems entirely from the GPU’s power cord, 
without requiring intimate knowledge of memory registers, 
without intricate signal capture, nor access to input-output variables. 
Furthermore, we optimize parameter assignment as an integer programming problem, 
allowing us to tackle arbitrary networks without choosing between parameter templates for hidden models. 
We validate across various GPUs and our extracted networks achieve transfer attack rates 
of $55.19-77.36$\% (VGG) $84.47-98.56$\% (ResNet), 
with existing works demonstrating $51.53$\% (VGG)~\cite{yu2020deepem} and $75.9$\% (ResNet)~\cite{hu2020deepsniffer}. 
Lastly, our study includes diverse activation functions, 
normalization layers, pooling variants, and recurrent layers unexplored in previous literature.

\paraspace
\paragraph{Limitations.}
In our formulation, we assume networks progress in a linear 
fashion and do not handle complex graph networks with intricate branching topologies.
We cannot tell if a network is trained with dropout 
since dropout layers do not appear at inference time.
Indeed, any operation that only appears during training 
is beyond the capability of magnetic side channel snooping.

It is an assumption of our method that the target network is observed during inference and not during training.
Our method may naturally extend to extracting within training phases 
by ignoring signals related to back-propagation and focusing only on the feed-forward steps. 
This extension would require the additional processing to isolate the forward pattern and sanitize
other signals auxiliary to the model architecture,
which we do not explore within the scope of this work.
Access to training would provide numerous example signatures of the network to work with, 
rather than the more restricted single source acquired during inference, 
which might in turn increase the recovery accuracy and robustness of our method. 
Given the ability to single out forward passes, 
one could straightforwardly apply our method to batches in every training epoch 
and consolidate the proposed networks using statistics and heuristics on the candidate models.
Our focus on inferences stems from our threat model, 
which does not require access to hardware during training.

\paraspace
\paragraph{Conclusion.}
We set out to study what can be learned from passively listening to a magnetic
side channel in the proximity of a running GPU.  
Our prototype shows it is
possible to extract both the high-level network topology and detailed hyperparameters.
To better understand the robustness and accuracy, we
collected a dataset of magnetic signals by inferencing through thousands of
layers on four different GPUs.  
We also investigated how one might use this
side channel information to turn a black-box attack into a white-box transfer attack.

\Urlmuskip=0mu plus 1mu\relax
\bibliographystyle{plain}
\bibliography{references}

\clearpage
\appendix

\begin{strip}
\vspace{-6mm}
\begin{center}
\Large
\textbf{Supplementary Document}\\ 
\smallskip
\textbf{Can one hear the shape of a neural network?:\\ Snooping the GPU via Magnetic Side Channel}
\medskip
\end{center}
\end{strip}

\setcounter{figure}{0}
\setcounter{table}{0}
\makeatletter 
\renewcommand{\thefigure}{S\@arabic\c@figure}
\renewcommand{\thetable}{S\@arabic\c@table}
\makeatother

\secprespace
\section{Additional Experiments}\label{app:exp}
\secprespace

\subsection{Reconstruction Quality on ImageNet}\label{app:imagenet}
We treat ResNet18 and ResNet50 for ImageNet classification as our black-box models, 
and reconstruct them from their magnetic signals. We then train those reconstructed
networks and compare their test accuracies with the original networks' performance.
Both the reconstructed and original networks are trained with the same training dataset
for the same number of epochs. 
The results are shown in \tabref{imagenet}, where we report both top-1 and top-5 classification
accuracies. In addition, we also report a KL-divergence measuring the difference between
the 1000-class image label distribution (over the entire ImageNet test dataset) predicted by the original network and that predicted
by the reconstructed network. Not only are those KL-divergence values small, we
also observe that for the reconstructed network that has a smaller KL-divergence from the original network (i.e., ResNet18),
its performance approaches more closely to the original network.


\begin{table}[h]
\centering
\caption{Model reconstruction evaluated on ImageNet classification.}\label{tab:imagenet}
\resizebox{\columnwidth}{!}{
 \begin{tabular}{l|cc|cc}
\bottomrule
\multirow{2}{*}{Model}
 & \multicolumn{2}{c|}{ResNet18} & \multicolumn{2}{c}{ResNet50} \\
 & Original & Extracted &Original & Extracted \\ \hline
 Top-1 Acc. &64.130 & 64.608  & 62.550 & 61.842\\
 Top-5 Acc. &86.136 & 86.195 & 85.482 & 84.738\\
 KL Div.   & - & 2.39 & - & 4.85\\
\toprule
 \end{tabular}
 }
\end{table}

\begin{table}[h]
\centering
\caption{\textbf{DNN estimation accuracies.}  
Using the 1804 convolutional layers in our test dataset,
we measure the accuracies of our DNN models for estimating the
convolutional layers' hyperparameters. Here, we break the accuracies down 
into the accuracies for individual hyperparameters.
}\label{tab:dnn}
\resizebox{\columnwidth}{!}{
 \begin{tabular}{r|ccccc}
\bottomrule
           & Kernel & Stride & Padding & Image-in & Image-out \\ \hline
 Precision &0.971 &0.976 &0.965 &0.968 &0.965 \\
 Recall    &0.969 &0.975 &0.964 &0.969 &0.968 \\
 F1 Score  &0.969 &0.975 &0.962 &0.967 &0.965 \\
\toprule
 \end{tabular}
 }
\end{table}

\begin{figure}[t]
    \centering
    \includegraphics[width=0.99\linewidth]{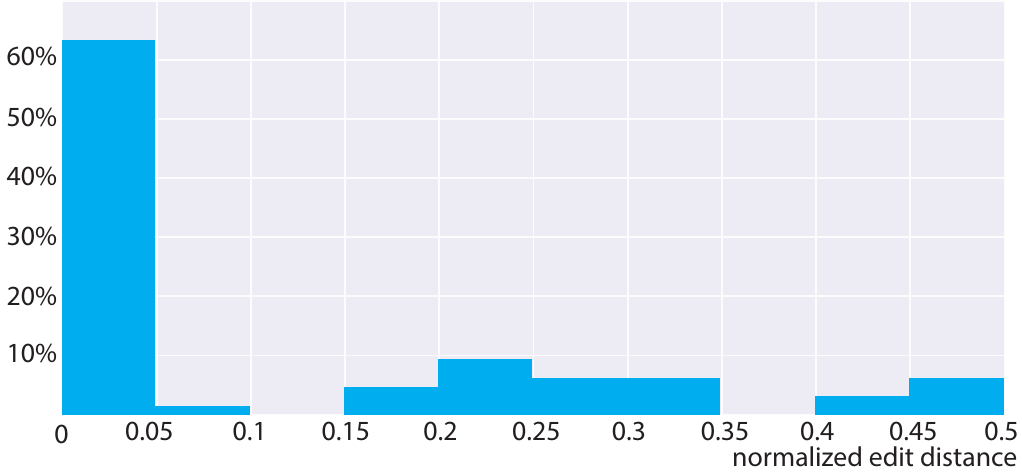}    
    \vspace{-2mm}
    \caption{\textbf{Distribution of normalized Levenshtein distance on dataset.}
    We plot the distribution of the normalized Levenshtein distances between
    the reconstructed and target networks. This results, corresponding to \tabref{classify1}
    in the main text, use signals collected on Nvidia Titan V.
    \label{fig:distancetitanv}}
\end{figure}

\begin{figure}[t]
    \centering
    \includegraphics[width=0.99\linewidth]{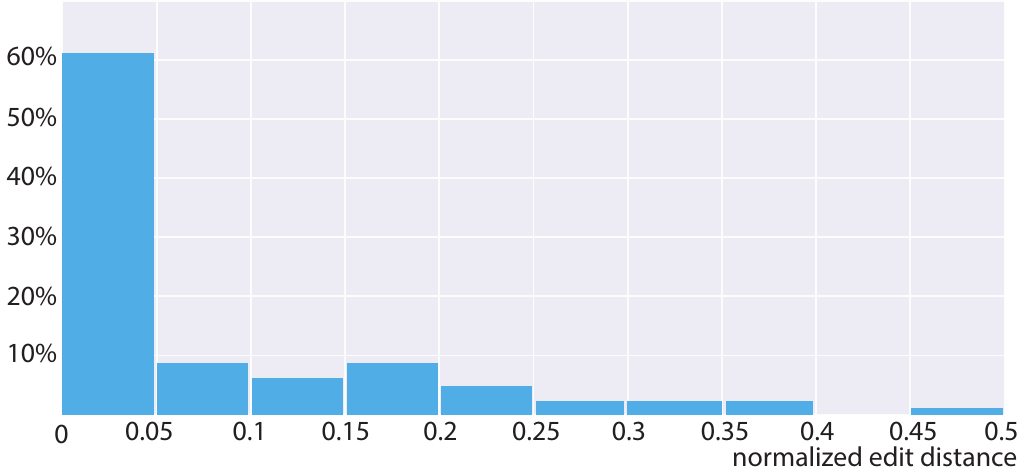}
    \vspace{-2mm}
    \caption{\textbf{Distribution of normalized Levenshtein distance across GPUs.}
    An experiment comparing the distribution of the normalized Levenshtein distances 
    between two Nvidia GTX-1080 GPUs.
    One is used for collecting training signals, and the other is used for 
    testing our side-channel-based reconstruction algorithm.
    \label{fig:distancegpu}}
\end{figure}



\secprespace
\section{Variations to our Approach}\label{app:variations}
\secpostspace

\subsection{Recovering Encoder-Decoder}
Encoder-decoder networks can similarly be recovered by our method. 
The handling of decoders differs in the constraints used 
for the integer programming problem to ensure parameter consistency. 
Decoders are composed of the same functional steps as encoders, 
but potentially grow in size across steps. 
Treating decoders requires reassessing the optimization constraints 
to enforce that layers remain the same size or grow. 
Recovering encoder-decoder networks introduces the additional challenge 
of pinpointing the switch from encoder to decoder. 
However, if this transition can be detected or approximated, 
an additional boundary constraint could be introduced to our 
integer programming formulation that would allow the network 
to be split into two optimization problems constrained to align at the transition. 
This formulation would provide a natural extension of our method 
to handle encoder-decoder networks, however any such attempts remains future work.

\subsection{Software Optimizations}
Machine learning frameworks sometimes allow for 
performance-tuned alternatives for frequently used operations. 
These inference optimizers have the effect of introducing new categories 
to the supervised BiLSTM classification task, 
either by merging commonly paired operations (i.e. fully connected step and batch norm) 
or introducing new operations altogether (i.e optimized mean computation for average pooling). 
Changing topology classification categories provides a defense 
against a BiLSTM classifier trained without optimizers in its dataset. 
However, there are no stipulations against building a dataset consisting of inferences 
where optimizers are both turned on and off, 
generating an encompassing BiLSTM classifier that would be robust to such defenses.

\end{document}